\documentclass{article}

\usepackage{amsmath}
\usepackage[pdftex]{graphicx}
\usepackage{mlspconf}

\usepackage[pdftex, hidelinks]{hyperref}

\usepackage[pdftex]{xcolor}
\usepackage[caption=false,font=footnotesize]{subfig}

\usepackage{cite, url}

\usepackage{array, amssymb, amsfonts, bm, mathtools}
\usepackage{booktabs}
\usepackage{algorithm, algpseudocode}
\usepackage{multirow}
\usepackage{xargs}
\usepackage{enumitem}

\usepackage{comment}
\usepackage{kantlipsum}

\newcommand{\setR}{\mathbb{R}}
\newcommand{\setRp}{\mathbb{R}_+}
\newcommand{\setC}{\mathbb{C}}

\newcommand{\T}{\mathsf{T}}
\newcommand{\adj}{\mathsf{H}}

\newcommand{\sep}[1][tfk]{g_{#1}}
\newcommand{\loc}[1][kd]{h_{#1}}

\NewDocumentCommand\newletter{m m o m m}{
\NewDocumentCommand#1{s t@ o}{%
\IfBooleanTF{##1}{\mathbf{\MakeUppercase{#2}}\IfValueT{#3}{^{#3}}}{%
\IfBooleanTF{##2}{\mathbf{#2}\IfValueT{#3}{^{#3}}_{\IfValueTF{##3}{##3}{#5}}}{%
{#2}\IfValueT{#3}{^{#3}}_{\IfValueTF{##3}{##3}{#4}}%
}}}}

\NewDocumentCommand\newletterbm{m m o m m}{
\NewDocumentCommand#1{s t@ o}{%
\IfBooleanTF{##1}{\bm{\MakeUppercase{#2}}\IfValueT{#3}{^{#3}}}{%
\IfBooleanTF{##2}{\bm{#2}\IfValueT{#3}{^{#3}}_{\IfValueTF{##3}{##3}{#5}}}{%
{#2}\IfValueT{#3}{^{#3}}_{\IfValueTF{##3}{##3}{#4}}%
}}}}

\newletter{\x}{x}{tfm}{tf}
\newletter{\s}{s}{tfk}{tf}
\newletter{\n}{n}{tfm}{tf}
\newletter{\sv}{a}{fdm}{fd}
\newletter{\svt}{b}{fdm}{fd}
\newletterbm{\psd}{\lambda}{tfk}{tf}
\newcommand{\scm}[1][fd]{\mathbf{H}_{#1}}
\newcommand{\scmt}[1][fd]{\mathbf{G}_{#1}}

\newletterbm{\avepwr}{\tilde{\lambda}}{}{}

\newletter{\z}{z}{tfk}{tf}
\newletter{\w}{w}{kd}{k}
\newletter{\ez}{\hat{z}}{tfk}{tf}
\newletter{\ew}{\hat{w}}{kd}{k}

\newletterbm{\zrate}{\pi}{tk}{t}
\newletterbm{\wrate}{\phi}{d}{}

\newcommand{\allparams}{\bm{\Theta}}

\newcommand{\elbo}{\mathcal{L}}

\newcommand{\distcmpnormal}[2]{\mathcal{N}_{\mathbb{C}}\left({#1}, {#2}\right)}
\newcommand{\distcmpinvwishart}[2]{\mathcal{IW}_{\mathbb{C}}\left({#1}, {#2}\right)}
\newcommand{\distcategorical}[1]{\mathrm{Cat}\left({#1}\right)}

\newcommand{\E}{\mathbb{E}}
\newcommand{\KL}{\mathbb{KL}}

\title{Deep Bayesian Unsupervised Source Separation \\ Based on a Complex Gaussian Mixture Model}

\name{Yoshiaki Bando$^1$, Yoko Sasaki$^1$, Kazuyoshi Yoshii$^2$\thanks{Thanks to JSPS KAKENHI 18H06490 and 19H04137 for funding.}}
\address{$^1$National Institute of Advanced Industrial Science and Technology, Japan, \\ $^2$RIKEN AIP / Kyoto University, Japan}

\begin{document}

\setlength\abovedisplayskip{1.0mm}
\setlength\belowdisplayskip{1.0mm}

\maketitle
\begin{abstract}
 This paper presents an unsupervised method that trains neural source separation by using only multichannel mixture signals.
 Conventional neural separation methods require a lot of supervised data to achieve excellent performance.
 Although multichannel methods based on spatial information can work without such training data,
   they are often sensitive to parameter initialization and degraded with the sources located close to each other.
 The proposed method uses a cost function based on a spatial model called a complex Gaussian mixture model (cGMM).
 This model has the time-frequency (TF) masks and direction of arrivals (DoAs) of sources as latent variables
 and is used for training separation and localization networks that respectively estimate these variables.
 This joint training solves the frequency permutation ambiguity of the spatial model in a unified deep Bayesian framework.
 In addition, the pre-trained network can be used not only for conducting monaural separation but also for efficiently initializing a multichannel separation algorithm.
 Experimental results with simulated speech mixtures showed that our method outperformed a conventional initialization method.
\end{abstract}
\begin{keywords}
 Unsupervised source separation, complex Gaussian mixture model, deep Bayesian learning
\end{keywords}


\section{Introduction}
Deep neural networks (DNNs) have demonstrated excellent performance in source separation tasks, such as speech separation~\cite{hershey2016deep,kolbaek2017multitalker,drude2018deep} and music separation~\cite{jansson2017singing,luo2017deep}.
Permutation invariant training (PIT), for example, trains a DNN to output time-frequency (TF) masks for corresponding sources.
Such a method requires a large number of clean source signals and their mixtures for supervised training.
It is, however, practically difficult to prepare such supervised data in several tasks.
Source separation for audio scene analysis, for example, has to separate daily-life audio events,
  which are generally captured only in mixture recordings.
This calls for an unsupervised method that works without any supervised data.

Unsupervised source separation based on spatial information observed in multichannel recordings has widely been studied~\cite{ozerov2010mnmf, kim2010realtime, ono2011stable,otsuka2014bayesian}.
A standard approach is to estimate TF masks from phase and power differences among microphones.
A complex Gaussian mixture model (cGMM)~\cite{higuchi2016robust,otsuka2014bayesian,azcarreta2018permutation}, for example, represents such spatial characteristics as spatial covariance matrices (SCMs)
  and estimates TF masks by clustering TF bins.
Since the cGMM is independently formulated at frequency bins, it has permutation ambiguity that the indices of sources are not aligned over frequency bins.
This ambiguity can be resolved by aligning estimated sources based on the direction of arrival (DoA) of each source,
 and several methods have been proposed to jointly estimate the TF masks and DoAs~\cite{otsuka2014bayesian,azcarreta2018permutation}.
The directional information also makes it possible to estimate the number of sources~\cite{otsuka2014bayesian,azcarreta2018permutation}, which many methods require in advance~\cite{ozerov2010mnmf, kim2010realtime, ono2011stable}.
The multichannel methods, however, are often sensitive to parameter initialization and degraded when the sources are located close to each other.

\begin{figure}[t]
 \centering
 \includegraphics[width=\hsize]{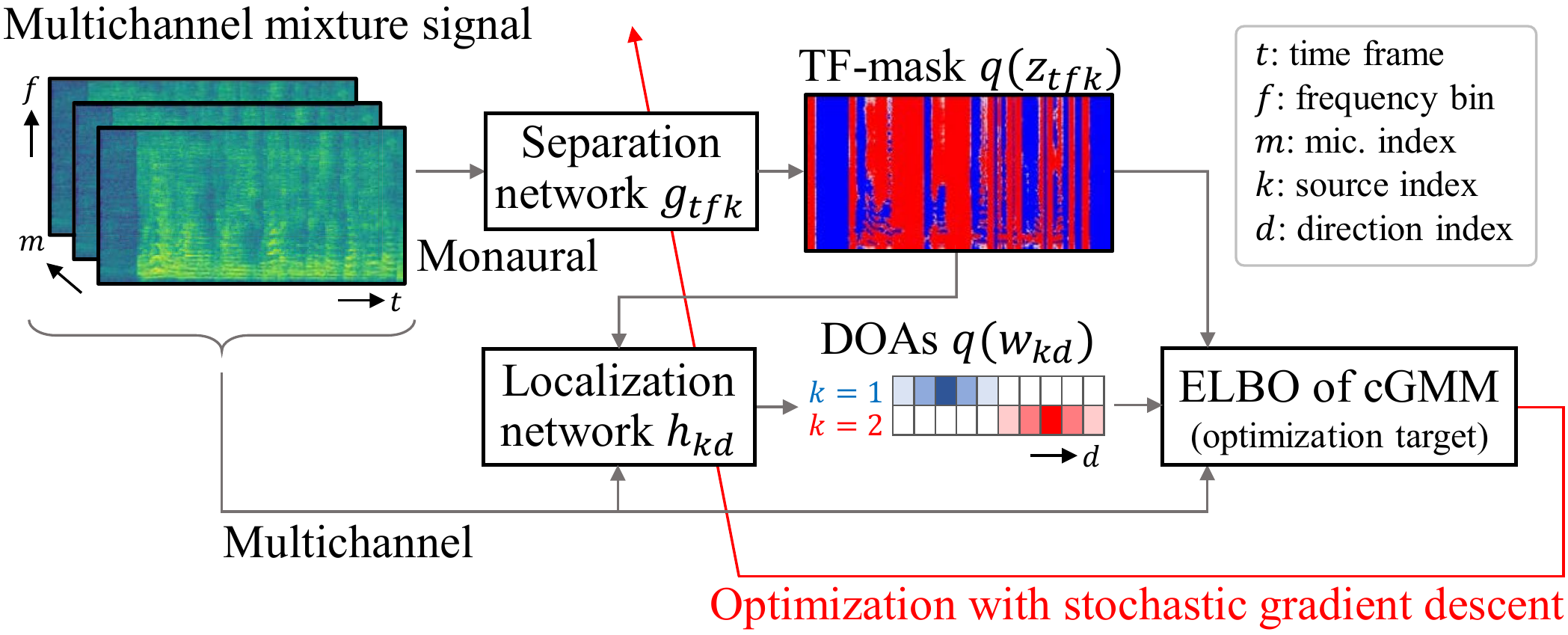}
 \vspace{-9mm}
 \caption{Overview of cGMM-based unsupervised training.}
 \label{fig:overview}
 \vspace{-3mm}
\end{figure}

Unsupervised training for neural source separation using multichannel mixture signals has recently gained a lot of attention~\cite{drude2019unsupervised,seetharaman2018bootstrapping,tzinis2018unsupervised,drude2019unsupervised2}.
One approach is to generate supervised data by using multichannel separation methods~\cite{drude2019unsupervised,seetharaman2018bootstrapping,tzinis2018unsupervised}.
This approach suffers from the estimation errors of the multichannel methods mentioned above.
To solve this problem, Drude et~al.~\cite{drude2019unsupervised2} trained a separation network by directly optimizing the likelihood function of a cGMM.
They reported that the performance of a conventional multichannel method was improved by initializing it with the network output.
To solve the frequency permutation ambiguity by using the correlation of TF masks over frequency bins~\cite{sawada2007measuring},
  the method requires the number of latent sources in advance.
It is thus difficult to apply this method for recordings of daily-life audio events, which include an unknown number of source signals.

To tackle this problem,
  we solve the frequency permutation ambiguity by jointly training separation and localization networks instead of using the correlation of masks (Fig.~\ref{fig:overview}).
The objective function is derived as an evidence lower bound (ELBO)~\cite{kingma2013auto} of a cGMM that has TF masks and DoAs as latent variables.
Given the geometry of a microphone array,
  the two networks are trained to respectively estimate the posterior probabilities of the TF masks and DoAs.
Since DoAs can be used for counting the number of sources in a mixture recording,
our framework could be extended to deal with training data including an unknown number of sources by utilizing a non-parametric Bayesian model~\cite{kurihara2007collapsed, otsuka2014bayesian}.

The main contribution of this paper is to resolve the frequency permutation ambiguity with a unified deep Bayesian framework during the unsupervised training.
We show that the separation network can be trained from random weights by maximizing the ELBO without any additional solvers or steps for the permutation problem.
The trained network can be used not only for conducting monaural source separation but also for efficiently initializing a multichannel separation algorithm.
Experimental results also show that the proposed method outperforms an existing initialization method.

\section{Related Work}
This section overviews cGMM-based TF clustering and then introduces unsupervised neural source separation.

\subsection{Complex Gaussian mixture models}
A popular approach to separating a multichannel mixture signal is to mask each TF bin~\cite{araki2009blind,mandel2007algorithm,higuchi2016robust,otsuka2014bayesian,azcarreta2018permutation}.
This mask is conventionally estimated by clustering hand-crafted features at each TF bin ~\cite{araki2009blind,mandel2007algorithm}.
To directly conduct a clustering on a multichannel spectrogram,
  probabilistic mixture models for a multichannel observation have been studied~\cite{higuchi2016robust,otsuka2014bayesian,azcarreta2018permutation}.
The cGMM, for example, represents the multichannel spectrogram as a mixture of complex Gaussian distributions with SCMs and power spectral densities of sources~\cite{higuchi2016robust}.
A complex angular central Gaussian mixture model (cACGMM)~\cite{ito2016complex} is defined on a multichannel spectrogram normalized by power at each TF bin.
It has been proven that the expectation-maximization (EM) algorithms for the cGMM and cACGMM are equivalent~\cite{ito2016complex}.
Since these models are independently formulated at frequency bins, they have the frequency permutation ambiguity.
To solve this problem, a cGMM-based method estimates the TF mask and DoA of each source by using an inverse Wishart mixture prior on the SCMs~\cite{azcarreta2018permutation}.
Wishart distributions of this mixture represent potential DoAs characterized by using premeasured steering vectors.
Another cGMM-like spatial model inspired by latent Dirichlet allocation (LDA)~\cite{otsuka2014bayesian} jointly estimates the TF mask and the DoA of each source, and the number of sources in a unified framework.
This joint estimation is conducted with a collapsed Gibbs sampling by assuming a hierarchical Dirichlet process.

\subsection{Unsupervised training of neural source separation}
Unsupervised training of neural source separation has been studied by using visual information~\cite{efrosaudio,rouditchenko2019self} or multichannel recordings~\cite{drude2019unsupervised,seetharaman2018bootstrapping,tzinis2018unsupervised}.
The audio-visual-based methods use video recordings that capture the audio events and corresponding visual events,
  such as music signals and corresponding performances~\cite{efrosaudio,rouditchenko2019self}.
These methods are based on the co-occurrence of the audio and visual events and train a network so that the separated signals correlate to the visual events.
Multichannel-audio-based methods, on the other hand, can train a DNN to separate sound sources out of view or behind obstacles.
Tzinis et~al.~\cite{tzinis2018unsupervised} trained a monaural separation network by using source signals estimated by applying $K$-means clustering on interchannel phase differences (IPDs) between two microphones.
Almost simultaneously, Drude et~al.~\cite{drude2019unsupervised} proposed a similar approach that uses signals separated by the cACGMM~\cite{ito2016complex}.
They reported that the cACGMM performance was improved by initializing it with the pre-trained separation network.
Seetharaman et~al.~\cite{seetharaman2018bootstrapping} designed a loss function weighted by a confidence measure of the estimated references.
Drude et~al.~\cite{drude2019unsupervised2} also proposed a novel approach that directly trains a separation network from the cACGMM likelihood.
They applied the method to noisy speech recordings and reported that the performance of automatic speech recognition was superior to that of their previous approach mentioned above.


\section{Deep Bayesian Source Separation}
\vspace{-0.5mm}
The proposed method trains separation and localization networks by using only multichannel mixture signals
  and resolves the frequency permutation ambiguity in a unified framework.
This training is based on the LDA model~\cite{otsuka2012bayesian, otsuka2014bayesian}, which has TF masks and DoAs of sources as latent variables.
The objective function is derived as an ELBO of the spatial model, which consists of an expectation of the likelihood function and a Kullback-Leibler (KL) divergence between the network outputs and their prior distributions.
Since the existing studies~\cite{otsuka2012bayesian, otsuka2014bayesian} only show Bayesian inference for the LDA model,
  we also describe an EM algorithm of the model and initialize it with the pre-trained network.
  
\subsection{Probabilistic generative model}
\vspace{-0.5mm}
To jointly estimate the TF-masks and DoAs  of latent sound sources,
 an observed $M$-channel spectrogram $\x@ \in \setC^M$ is represented as a sum of $K$ source spectrograms $\s \in \setC$:
\begin{align}
 \x@ = \sum_{k=1}^K \sum_{d=1}^D \z \w \left( \sv@ \s \right), \label{eq:lda-mix}
\end{align}
where $\z \in \{0, 1\}$ ($\sum_{k=1}^K \z = 1$) is a TF mask that indicates which source is relevant at each TF bin,
  $\w \in \{0, 1\}$ ($\sum_{d=1}^D \w = 1$) is a DoA variable that assigns source $k$ to a DoA candidate $d \in \{1, \ldots, D\}$,
  and $\sv@ \in \setC^M$ is a steering vector for direction $d$.
As in other cGMMs~\cite{higuchi2016robust,otsuka2014bayesian,azcarreta2018permutation}, the TF mask $\z$ is introduced by assuming a sparseness that each TF bin has exclusively one relevant source.
The potential directions $d$ are, in this paper, assumed as directions with an angular interval of 5$^\circ$ on a horizontal plane ($D=72$).

The TF masks and DoAs are estimated as their posterior probabilities by putting prior distributions on them.
Since the activity of each source changes over time frames,
  a frame-wise categorical distribution (denoted as $\mathrm{Cat}$) is put on the TF-masks $z_{tfk}$ as follows:
\begin{align}
 [\z[tf1], \ldots, \z[tfK]]^\T \mid \zrate@ \sim \distcategorical{\zrate[t1], \ldots, \zrate[tK]},
\end{align}
where $\pi_{tk} \in \mathbb{R}_+$ ($\sum_{k=1}^K \pi_{tk} = 1$) is a model parameter to be estimated.
On the other hand, the following categorical distribution is put on $\w$ as follows:
\begin{align}
  \w@ = [\w[k1], \ldots, \w[kD]]^\T \sim \distcategorical{\wrate[1], \ldots, \wrate[D]}. \label{eq:pphi}
\end{align}
where $\wrate \in \setRp$ ($\sum_{d=1}^D \wrate = 1$) is a model parameter.

Each source spectrogram $\s$ is assumed to follow a zero-mean complex Gaussian distribution:
\begin{align}
  \s \sim \distcmpnormal{0}{\psd}, \label{eq:s}
\end{align}
where $\distcmpnormal{\mu}{\sigma^2}$ is a complex Gaussian distribution with mean $\mu$ and variance $\sigma^2$, and $\psd \in \setRp$ represents the power spectral density of source $k$.
Using \eqref{eq:lda-mix} and \eqref{eq:s}, an observed mixture signal $\x@$ is found to follow a multivariate complex Gaussian mixture distribution as follows:
\begin{align}
  \x@ \sim \prod_{k=1}^K \prod_{d=1}^{D} \distcmpnormal{\bm{0}}{\psd \scm}^{\z\w},
\end{align}
where $\scm = \E[\sv@ \sv@^\adj] \in \setC^{M\times M}$ is a SCM of direction $d$.
To estimate $\scm$ while constraining it to direction $d$, the following complex inverse Wishart distribution is put on $\scm$:
\begin{align}
  \scm \sim \distcmpinvwishart{\nu}{(\nu - M)\scmt}, \label{eq:pw}
\end{align}
where $\mathcal{IW}_\mathbb{C}(\nu, \mathbf{G}) \propto |\mathbf{H}|^{-(\nu+{M})}\exp[-\mathrm{tr}(\mathbf{G} \mathbf{H}^{-1})]$ represents the complex inverse Wishart distribution, $\nu > M$ is a hyperparameter,
  $\scmt = \svt@ \svt@^\adj + \epsilon \mathrm{I} \in \setC^{M\times M}$ is a template SCM for direction $d$.
The $\svt@$ is a template steering vector for direction $d$ and prepared in advance, and $\epsilon\mathbf{I}$ ($\epsilon > 0$) is added to make $\scmt$ positive definite.

\vspace{-1.0mm}
\subsection{Variational inference framework}
\vspace{-1.0mm}
Both the proposed unsupervised training and multichannel separation are based on a variational inference that estimates the posterior distribution $p(\z*,\w*|\x*, \allparams)$,
 where $\allparams = \{ \scm[], \psd*, \zrate*, \wrate* \}$ represents the parameters obtained by point estimation.
Since it is difficult to analytically calculate the true posterior distribution $p(\z*, \w* |\x*, \allparams)$,
 we approximate it with the following variational posterior distribution:
\begin{align}
 p(\z*, \w*|\x*, \allparams) \approx q(\z*)q(\w*).
\end{align}
The variational inference is conducted by maximizing the following lower bound of the log marginal likelihood $p(\x*|\allparams)$:
\begin{align}
 \hspace{-2mm}\elbo &= \E_q \left[ \log p(\x* \mid \psd*, \scm[], \z*, \w*)\right] \nonumber \\
 &\hspace{3mm}- \KL\left[ q(\z*) | p(\z*|\zrate*)\right] - \KL\left[ q(\w*) | p(\w*|\wrate*)\right]. \label{eq:elbo}
\end{align}
The lower bound $\elbo$ is called an ELBO, and its maximization corresponds to the minimization of KL divergence between the variational and true posterior distributions.
This framework iteratively and alternately updates the variational posteriors $q$ and parameters $\allparams$ until convergence.

The SCM $\scm[]$ is updated with maximum a posteriori (MAP) estimation and the other parameters $\psd*$, $\zrate*$, and $\wrate*$ are updated with maximum likelihood estimation.
Since it is also difficult to analytically calculate these variables, we update them by using the ELBO \eqref{eq:elbo} as follows:
\begin{align}
 \scm &\leftarrow \frac{\scmt + \sum_{t,k=1}^{T,K} \ez\ew \frac{1}{\psd} \x@ \x@^\adj}{\nu + \sum_{t,k=1}^{T,K} \ez\ew + M}, \label{eq:m-step0} \\
 \psd &\leftarrow \frac{1}{M} \sum_{d=1}^D \ew \x@^\adj \scm^{-1} \x@,\\
 \zrate &\leftarrow \frac{1}{F}\sum_{f=1}^F \ez, \hspace{5mm} \wrate \leftarrow \frac{1}{K} \sum_{k=1}^K \ew, \label{eq:m-step}
\end{align}
where $\ez$ is $q(\z=1)$ and $\ew$ is $q(\w=1)$.

\subsection{Training based on amortized variational inference}
By using $N$ mixture signals $\x@^{(n)}$, we train separation and localization networks that respectively estimate the TF mask $z_{tfk}$ and DoA $w_{kd}$ (Fig.~\ref{fig:overview}).
The suffix $^{(n)}$ is hereinafter omitted
  because the objective function is a sum of the local loss value for each mixture signal $\x@^{(n)}$.
The separation network (denoted by $g_{tfk}$) takes as input a monaural log-magnitude spectrogram and expects the posterior distribution of the TF mask $q_g(\z=1)$:
\begin{align}
  q_g(\z=1) &= \ez = g_{tfk}(\log|\x*|), 
\end{align}
where $\log|\x*| \in \setR^{T\times F}$ denotes a monaural log-magnitude spectrogram.
We simply take the recording of the first microphone ($m=1$) as the input.
The localization network (denoted by $h_{kd}$), on the other hand, expects the probability that direction $d$ is selected for the $k$-th source $q_h(\w=1)$:
\begin{align}
  q_h(\w=1) &= \ew = h_{kd}\left( \bm{\omega} \right),
\end{align}
where $\bm{\omega} =\{ \omega_{kd}\}_{k,d=1}^{K,D}\in \setR^{K\times D}$ is an input feature that represents spatial characteristics.
Since it is difficult for networks to directly take complex numbers as input,
  we alternatively use the following Gaussian-mixture log likelihood:
\begin{align}
 \omega_{kd} &= \sum_{t=1}^{T}\sum_{f=1}^{F} \ez \log \distcmpnormal{\x@; \bm{0}}{\scmt}.
\end{align}

The training of networks $\sep$ and $\loc$ is conducted by maximizing the ELBO $\elbo$ for each mixture signal in the training data.
For numerical stability,
  we fix $\psd$ and $\scm$ to $\avepwr=\frac{1}{TFM}\sum_{t,f=1}^{T,F}\x@^\adj\x@$ and $\scmt$, respectively.
More specifically, the proposed training is conducted by iteratively executing the following three steps:
{\vspace{-2mm}%
\setlength{\leftmargini}{13pt}%
\begin{enumerate}
 \renewcommand{\labelenumi}{\arabic{enumi})}
 \setlength{\itemsep}{0pt}
 \setlength{\parskip}{0pt}
 \item predict TF masks $\ez$ and DoAs $\ew$ with $g_{tfk}$ and $h_{kd}$ for each mixture recording in a mini-batch,
 \item update model parameters $\allparams = \{ \zrate*, \wrate* \}$ with \eqref{eq:m-step}, and
 \item calculate $\elbo$ and update the network parameters by using a stochastic gradient descent (SGD) method.
\end{enumerate}%
\vspace{-2mm}}%
\noindent The ELBO $\elbo$ can be calculated as follows:
\begin{align}
 \hspace{-2mm}\elbo &= -\sum_{t,f,k,d=1}^{T,F,K,D} \ez \ew \left( \log|\scmt| + \frac{1}{\avepwr}\x@^\adj \scmt^{-1}\x@ \right) \nonumber \\
 &\hspace{-2mm}+ \sum_{t,f,k=1}^{T,F,K} \ez \log \frac{\zrate}{\ez} + \sum_{k,d=1}^{K,D} \ew \log \frac{\wrate}{\ew} + \mathrm{const}..
\end{align}
The loss value for a mini-batch is a sum of this local ELBO normalized with $\frac{1}{TF}$.
Our method trains neural networks to estimate posterior distributions for unseen observed data by using a training data prepared in advance.
This kind of training is called amortized variational inference~\cite{ranganath2014black, kingma2013auto}.

\subsection{Multichannel separation based on an EM-algorithm}
Although the trained network $\sep$ can be used to separate sources from a monaural mixture signal,
  it can also improve the performance of a multichannel EM algorithm by initializing TF masks with the network output.
The EM algorithm for the cGMM (EM-cGMM) alternately iterates the following E-step and M-step.
The E-step updates the TF masks $\ez$ and DoAs $\ew$  so that the ELBO $\elbo$ is maximized:
\newcommand{\uz}{\zrate \prod_{d=1}^D \distcmpnormal{\x@; \bm{0}}{\psd \scm}^{\ew}}
\newcommand{\uw}{\wrate \prod_{t,f=1}^{T,F} \distcmpnormal{\x@; \bm{0}}{\psd \scm}^{\ez}}
\begin{align}
 \ez &\leftarrow \frac{\uz}{\sum_{K=1}^K \uz}, \\
 \ew &\leftarrow \frac{\uw}{\sum_{d=1}^D \uw}.
\end{align}
The M-step, on the other hand, updates the parameters $\allparams$ by using \eqref{eq:m-step0}--\eqref{eq:m-step}.
Since the EM algorithm alternately updates these variables until convergence,
  the careful initialization is important to avoid falling into a local optimum.

The TF masks $\ez$ are initialized by using the output of the separation network $\sep$.
Since the localization network $\sep$ can potentially overfit to the spatial bias of the training data,
  we initialize the DoA $\ew$ by using the following formula instead of the output of $\loc$:
\begin{align}
 \ew \propto \exp \left( -\sum_{t=1}^T\sum_{f=1}^{F} \ez \x@^\adj \scmt^{-1}\x@ \right). \label{eq:init-w}
\end{align}

\section{Experimental Evaluation}
We conducted an evaluation with speech mixture signals generated by using simulated room impulse responses (RIRs).

\subsection{Dataset}
The mixture signals used in this evaluation were generated by convolving RIRs to source signals in the WSJ0-mix dataset~\cite{hershey2016deep},
  which is widely used for neural speech separation~\cite{hershey2016deep,drude2018deep,kolbaek2017multitalker}.
Each of the mixture signals in this dataset included two utterances from two randomly selected speakers in the WSJ0 corpus.
The two speech signals were mixed with a signal-to-noise ratio randomly chosen between $-5$ and $+5$\,dB.
The RIRs applied to the speech signals were simulated by using the image method\footnote{\url{https://github.com/ty274/rir-generator}}~\cite{allen1979image} with the room configuration randomly changed at each mixture signal between 5\,m$\times$5\,m$\times$3\,m and 10\,m$\times$10\,m$\times$4\,m.
We assumed a 4-channel microphone array with the diameter of 8\,cm located at the center of the room.
The source locations of two speech signals were randomly placed in the room.
The reverberation time (RT$_{60}$) was chosen at random between 0.2 and 0.4\,s.
The training and validation sets had 20,000 and 5,000 mixture signals, respectively.
The test set had 3,000 mixture signals whose speakers were separated from the training and validation sets.
We generated these signals with a sampling rate of 8\,kHz to reduce computational and memory costs.

\subsection{Experimental Condition}
The network architectures for the proposed method were experimentally determined as follows.
The separation network $\sep$ had three layers of bi-directional long short-term memory (BiLSTM),
  each with 600 units for each direction,
  and one fully connected layer followed by a softmax activation.
To reduce the parameters of the localization network $\loc$,
  the network $\loc$ consisted of three layers of 1D-convolution with the direction axis $d$ as the convolution axis of each layer.
The filter size of the convolution layers and the number of the filters were respectively set to 1 and 2 ($=K$).
The network $\loc$ outputs $\log \ew$ through a residual connection with the network input.

The separation network $\sep$ and localization network $\loc$ were jointly optimized
  using the Adam optimizer~\cite{kingma2014adam}.
The learning rate of the optimizer was initialized to $1.0 \times 10^{-3}$ and scaled down by 0.7 when the training loss value increased compared to that of the last epoch.
The spectrograms $\x@$ were obtained with the short-time Fourier transform (STFT) with a window length of 512 samples and a shifting interval of 128 samples.
The hyperparameters $\nu$ and $\epsilon$ were set to $M + 5.0$ and $1.0 \times 10^{-2}$, respectively.
The template steering vectors $\svt@$ were theoretically calculated under the planewave assumption.
Note that the $\svt@$ and the RIRs used for generating the mixture signals were much different because the sound sources were randomly located on the room under reverberant conditions.
We iterated the EM-cGMM 50 times.
The source signals were obtained by masking the observation $\x@$ with the estimated TF mask $\ez$.

The proposed method was compared with an independent vector analysis (AuxIVA)~\cite{ono2011stable}, and the supervised methods of PIT and deep clustering (DPCL)~\cite{hershey2016deep}.
AuxIVA was evaluated with two channels in all the four channels because it assumes that the number of microphones equals that of sources.
To use all the four microphones, we also evaluated an extension of AuxIVA (AuxIVA+) that conducts AuxIVA with a 4-channel input and clusters the separated signals into two sources~\cite{kitamura2015relaxation}.
The dimension of the latent space for DPCL was set to 20.
The separation networks for PIT and DPCL had the same condition as $\sep$ in the proposed method.
We compared the proposed neural initialization for EM-cGMM with the initialization method proposed by Otsuka et al.~\cite{otsuka2012bayesian,otsuka2014bayesian}.
Given a sufficient number of source classes $K$,
 this method splits directions $d=1, \ldots, D$ into $K$ groups and initializes the TF masks $\ez$ by using the directional information:
\begin{align}
 \ew &\propto \left\{ \begin{array}{ll}
	      1 & (k-1)\frac{D}{K} \leq d < k\frac{D}{K} \\
	      0 & \text{otherwise}
	     \end{array} \right., \label{eq:dini-w} \\
 \ez &\propto \exp\left( - \sum_{d=1}^D \ew \x@^\adj \scmt^{-1} \x@ \right). \label{eq:dini-z}
\end{align}
We set the number of source classes $K=6$ for this method.
The separation performance was evaluated using the signal-to-distortion ratio (SDR)~\cite{BSSEVAL}.

\subsection{Experimental Results}
\begin{table}
 \setlength\abovecaptionskip{0mm}
 \setlength\belowcaptionskip{1mm}
 
 \centering

 \caption{Averages and standard deviations of SDRs}
 \label{tab:sdr}
 
 \newcommand{\gr}{\color{gray}}
 \footnotesize
 \begin{tabular}{lc|cc|c}
  \toprule
  \multirow{2}{*}{Method} & \multirow{2}{*}{Init.} & \multicolumn{2}{c|}{\# of mics. $M$} & SDR \\
                          &                        & train & test                         & [dB] \\
  \midrule
  EM-cGMM & $\sep$ & 4 & 4 & \bf 10.6 $\pm$ 4.2 \\
  EM-cGMM & \eqref{eq:dini-w}--\eqref{eq:dini-z} & -- & 4 & 9.7 $\pm$ 5.0 \\
  \midrule
  AVI-cGMM & -- &  4 & 1 & 5.3 $\pm$ 4.5 \\
  \midrule
  AuxIVA+  & -- & -- & 4 & 9.9 $\pm$ 4.4 \\
  AuxIVA   & -- & -- & 2 & 5.6 $\pm$ 4.0 \\
  \midrule
  \gr PIT  & \gr -- & \gr 1 & \gr 1 & \gr 7.7 $\pm$ 4.5 \\
  \gr DPCL & \gr -- & \gr 1 & \gr 1 & \gr 6.9 $\pm$ 4.7 \\
  \bottomrule
 \end{tabular}
\end{table}

\begin{figure}[t]
 \centering
 \vspace{2mm}
 \includegraphics[width=1.0\hsize]{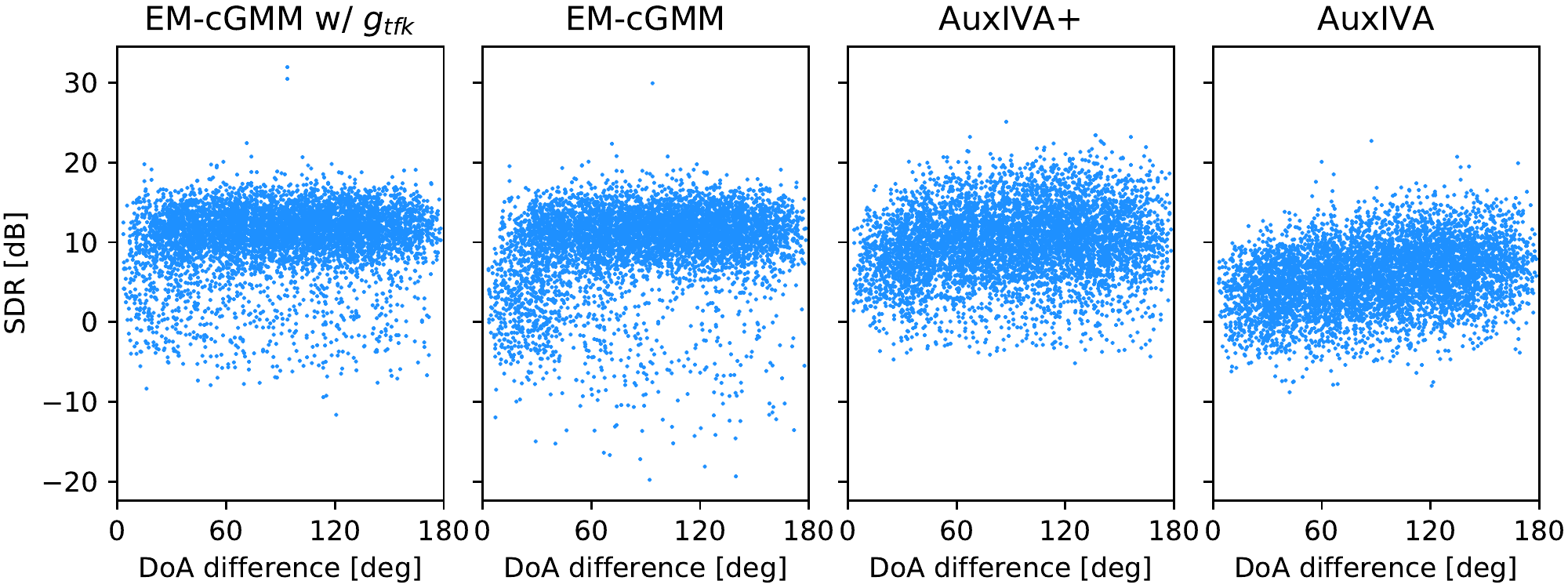}
 \vspace{-8mm}
 \caption{Scatter plots for the DOA difference of two sources and the corresponding SDR performance.}
 \label{fig:scat}
 \vspace{-6mm}
\end{figure}

The average SDRs for the test set were summarized in Table~\ref{tab:sdr}.
The EM-cGMM initialized with $\sep$ outperformed that with the conventional initialization (\eqref{eq:dini-w}--\eqref{eq:dini-z}).
In addition, it outperformed AuxIVA+, which uses the same number of microphones as the EM-cGMMs.
Fig~\ref{fig:scat} shows the relationship between the DoA differences and SDRs.
The EM-cGMM initialized with \eqref{eq:dini-w}--\eqref{eq:dini-z} significantly deteriorated when the DoA difference was less than 60$^\circ$.
The EM-cGMM initialized with $\sep$ improved the SDRs in such a condition.
The monaural separation with $\sep$ (AVI-cGMM) achieved 5.3\,dB in the average SDR.
When the mixture signals had speakers of difference genders (m+f in Table~\ref{tab:sdr-gender}),
  AVI-cGMM outperformed AuxIVA with 2-ch observations.

The initialization with $\sep$ occasionally decreased the performance regardless of the DoA differences,
  which is shown as the SDR results around $0$\,dB in Fig.~\ref{fig:scat}.
This is because $\sep$ (AVI-cGMM) deteriorated with the mixture signals of the same gender speakers (m+m and f+f in Table~\ref{tab:sdr-gender}), which are difficult to separate from spectral features.
Since the performances of PIT and DPCL were higher than that of the AVI-cGMM, 
  the $\sep$ has a potential to separate such signals.
Comparing AVI-cGMM with EM-cGMM initialized with $\sep$, 
  AVI-cGMM could be further improved by making it possible to estimate $\psd$ and $\scm$ during the training.
This extension will compensate with the mismatch between the fixed parameters $\avepwr$ and $\scmt$ and the observation due to reverberations and reflections.

\begin{table}
 \setlength\abovecaptionskip{0mm}
 \setlength\belowcaptionskip{1.5mm}

 \centering
 \caption{\fussy SDRs [dB] averaged by the genders (m: male, f: female) of the speakers in mixture signals.}
 \label{tab:sdr-gender}

 \newcommand{\gr}{\color{gray}}

 \footnotesize
 \begin{tabular}{lc|ccc}
  \toprule
  Method & Init. & m+m & f+f & m+f \\
  \midrule
  EM-cGMM  & $\sep$                               &  9.2 $\pm$ 4.9 & \bf 10.2 $\pm$ 5.2 & \bf 11.5 $\pm$ 3.1 \\
  EM-cGMM  & \eqref{eq:dini-w}--\eqref{eq:dini-z} &  9.5 $\pm$ 4.6 & 10.1 $\pm$ 5.2 &  9.7 $\pm$ 5.1 \\
  \midrule
  AVI-cGMM & -- &  2.0 $\pm$ 4.0 &  3.0 $\pm$ 4.4 &  7.9 $\pm$ 2.8 \\
  \midrule
  AuxIVA+  & -- & \bf 10.2 $\pm$ 4.4 &  9.3 $\pm$ 4.5 &  9.9 $\pm$ 4.4 \\
  AuxIVA   & -- &  5.7 $\pm$ 3.9 &  5.4 $\pm$ 4.1 &  5.7 $\pm$ 4.0 \\
  \midrule
  \gr PIT      & -- & \gr  4.9 $\pm$ 4.4 & \gr  5.2 $\pm$ 4.8 & \gr 10.1 $\pm$ 2.7 \\
  \gr DPCL     & -- & \gr  3.8 $\pm$ 4.6 & \gr  4.0 $\pm$ 4.8 & \gr  9.6 $\pm$ 2.8 \\
  \bottomrule
 \end{tabular}
 \vspace{-2mm}
\end{table}

\section{Conclusion}
We presented an unsupervised method that trains neural source separation by using only multichannel mixture signals.
The proposed method trains separation and localization networks by using a cost function based on a cGMM that has the TF masks and DoAs as latent variables.
This joint training enables us to resolve the frequency permutation ambiguity without any additional solvers or steps.
In addition, the trained network can also be used for efficiently initializing the cGMM-based multichannel EM algorithm.
We experimentally confirmed that the proposed initialization method outperformed a conventional initialization method.
To deal with the training data having an unknown number of sources,
   we plan to train a separation network while estimating the number of sources with the directional information.
We also plan to improve the proposed training method with the joint estimation of SCMs and power spectral densities.

\fussy
\let\oldthebibliography\thebibliography
\let\endoldthebibliography\endthebibliography
\renewenvironment{thebibliography}[1]{
  \begin{oldthebibliography}{#1}
    \setlength{\itemsep}{0.2em}
}
{
  \end{oldthebibliography}
}
\bibliographystyle{IEEEtran}
\bibliography{reference}

\end{document}